\documentclass[twocolumn,aps,prl,superscriptaddress,showpacs]{revtex4-1}
\usepackage{graphicx}
\usepackage{amsmath}
\usepackage{epsfig}
\usepackage[utf8]{inputenc}

\begin{document}
\title{Device calibration impacts security of quantum key distribution}

\author{Nitin Jain}
\email{nitin.jain@mpl.mpg.de}
\affiliation{Max Planck Institute for the Science of Light, G\"{u}nther-Scharowsky-Str.\ 1, Bau 24, 91058 Erlangen, Germany}
\affiliation{Institut f\"{u}r Optik, Information und Photonik, University of Erlangen-Nuremberg, Staudtstra\ss e 7/B2, 91058 Erlangen, Germany}

\author{Christoffer Wittmann}
\affiliation{Max Planck Institute for the Science of Light, G\"{u}nther-Scharowsky-Str.\ 1, Bau 24, 91058 Erlangen, Germany}
\affiliation{Institut f\"{u}r Optik, Information und Photonik, University of Erlangen-Nuremberg, Staudtstra\ss e 7/B2, 91058 Erlangen, Germany}

\author{Lars Lydersen}
\affiliation{Department of Electronics and Telecommunications, Norwegian University of Science and Technology, NO-7491 Trondheim, Norway}
\affiliation{University Graduate Center, NO-2027 Kjeller, Norway}

\author{Carlos Wiechers}
\affiliation{Max Planck Institute for the Science of Light, G\"{u}nther-Scharowsky-Str.\ 1, Bau 24, 91058 Erlangen, Germany}
\affiliation{Institut f\"{u}r Optik, Information und Photonik, University of Erlangen-Nuremberg, Staudtstra\ss e 7/B2, 91058 Erlangen, Germany}
\affiliation{Departamento de F\'isica, Campus Le\'on, Universidad de Guanajuato, Lomas del Bosque 103, Fracc.\ Lomas del Campestre, 37150, Le\'on, Gto, M\'exico}

\author{Dominique Elser}
\affiliation{Max Planck Institute for the Science of Light, G\"{u}nther-Scharowsky-Str.\ 1, Bau 24, 91058 Erlangen, Germany}
\affiliation{Institut f\"{u}r Optik, Information und Photonik, University of Erlangen-Nuremberg, Staudtstra\ss e 7/B2, 91058 Erlangen, Germany}

\author{Christoph Marquardt}
\affiliation{Max Planck Institute for the Science of Light, G\"{u}nther-Scharowsky-Str.\ 1, Bau 24, 91058 Erlangen, Germany}
\affiliation{Institut f\"{u}r Optik, Information und Photonik, University of Erlangen-Nuremberg, Staudtstra\ss e 7/B2, 91058 Erlangen, Germany}

\author{Vadim Makarov}
\affiliation{Department of Electronics and Telecommunications, Norwegian University of Science and Technology, NO-7491 Trondheim, Norway}
\affiliation{University Graduate Center, NO-2027 Kjeller, Norway}

\author{Gerd Leuchs}
\affiliation{Max Planck Institute for the Science of Light, G\"{u}nther-Scharowsky-Str.\ 1, Bau 24, 91058 Erlangen, Germany}
\affiliation{Institut f\"{u}r Optik, Information und Photonik, University of Erlangen-Nuremberg, Staudtstra\ss e 7/B2, 91058 Erlangen, Germany}

\date{\today}

\begin{abstract}
Characterizing the physical channel and calibrating the cryptosystem hardware are prerequisites for establishing a quantum channel for quantum key distribution (QKD). Moreover, an inappropriately implemented calibration routine can open a fatal security loophole. We propose and experimentally demonstrate a method to induce a large temporal detector efficiency mismatch in a commercial QKD system by deceiving a channel length calibration routine. We then devise an optimal and realistic strategy using faked states to break the security of the cryptosystem. A fix for this loophole is also suggested.
\end{abstract}


\maketitle

Quantum key distribution~(QKD) offers unconditionally secure communication as eavesdropping disturbs the transmitted quantum states, which in principle leads to the discovery of the eavesdropper Eve~\cite{qc}. However, practical QKD implementations may suffer from technological and protocol-operational imperfections that Eve could exploit in order to remain concealed~\cite{revw1,blckppr}. 

Until now, a variety of eavesdropping strategies have utilized differences between the theoretical model and the practical implementation, arising from (technical) imperfections or deficiencies of the components. Ranging from photon number splitting and Trojan-horse, to leakage of information in a side channel, time-shifting and phase-remapping, several attacks have been proposed and experimentally demonstrated~\cite{pns,trojanH,sidechans07,zhao08,phrmp10}. Recently, proof-of-principle attacks~\cite{larsnp10,carlosNlars10,gerhardt10} based on the concept of faked states~\cite{makarov05} have been presented. Eve targets imperfections of avalanche photodiode (APD) based single-photon detectors~\cite{props} that allow her to control them remotely.  

Another important aspect of QKD security not yet investigated, however, is the calibration of the devices. A QKD protocol requires a classical and a quantum channel; while the former must be authenticated, the latter is merely required to preserve certain properties of the quantum signals~\cite{revw1,revw2}. The establishment of the quantum channel remains an implicit assumption in security proofs: channel characterization (e.g.\ channel length) and calibration of the cryptosystem hardware, especially the steps involving two-party communication, haven't yet been taken into account. As we show, the calibration of the QKD devices must be carefully implemented, otherwise it is prone to hacks that may strengthen existing, or create new eavesdropping opportunities for Eve. 

\begin{figure}
\centerline{\includegraphics[width=6.7cm]{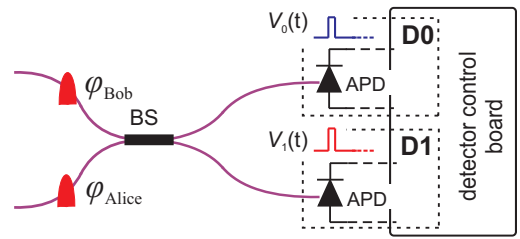}}
\caption{Typical detection system in a Mach-Zehnder interferometer based QKD implementation: The bit and basis choices of Alice and Bob (phases $\varphi_{\rm Alice}$ and $\varphi_{\rm Bob}$) determine the interference result at the 50:50 beam splitter (BS), or which of the two detectors D0 or D1 would click. It is thus crucial that D0 and D1 are indistinguishable to the outside world (i.e.\ Eve). If gated mode APDs are employed, the detector control board ensures that the activation of D0 and D1 (via voltage pulses $V_0(t)$ and $V_1(t)$) happens almost simultaneously, to nullify any existing temporal efficiency mismatch.}
\label{dsb}
\end{figure}

In this Letter, we propose and experimentally demonstrate the hacking of a vital calibration sequence during the establishment of the quantum channel in the commercial QKD system Clavis2 from ID~Quantique~\cite{clavis2guide}. Eve induces a parameter mismatch~\cite{makarov06} between the detectors that can break the security of the QKD system. Specifically, she causes a temporal separation of the order of $450$ ps of the detection efficiencies by deceiving the detection system, shown in Fig.~\ref{dsb}. This allows her to control Bob's detection outcomes using time, a parameter already shown to be instrumental in applying a time-shift attack (TSA) \cite{zhao08}. Alternatively, she could launch a faked-state attack (FSA) \cite{makarov06} for which we calculate the quantum bit error rate (QBER) under realistic conditions. Since FSA is an intercept-resend attack, Eve has full information-theoretic knowledge about the key as long as Alice and Bob accept the QBER at the given channel transmission $T$, and do not abort key generation~\cite{gllp}. Constricting our FSA to match the raw key rate expected by Bob and Alice, i.e.\ maintaining $T$ at nearly the exact pre-attack level, we find that the security of the system is fully compromised. Our hack has wide implications: most practical QKD schemes based on gated APDs, in both plug-and-play and one-way configurations~\cite{pnp,pnpvar,onewayqkd}, need to perform channel characterization and hardware calibration regularly. A careful implementation of these steps is required to avoid leaving inadvertent backdoors for Eve. 

\begin{figure}
\centerline{\includegraphics[width=8.5cm]{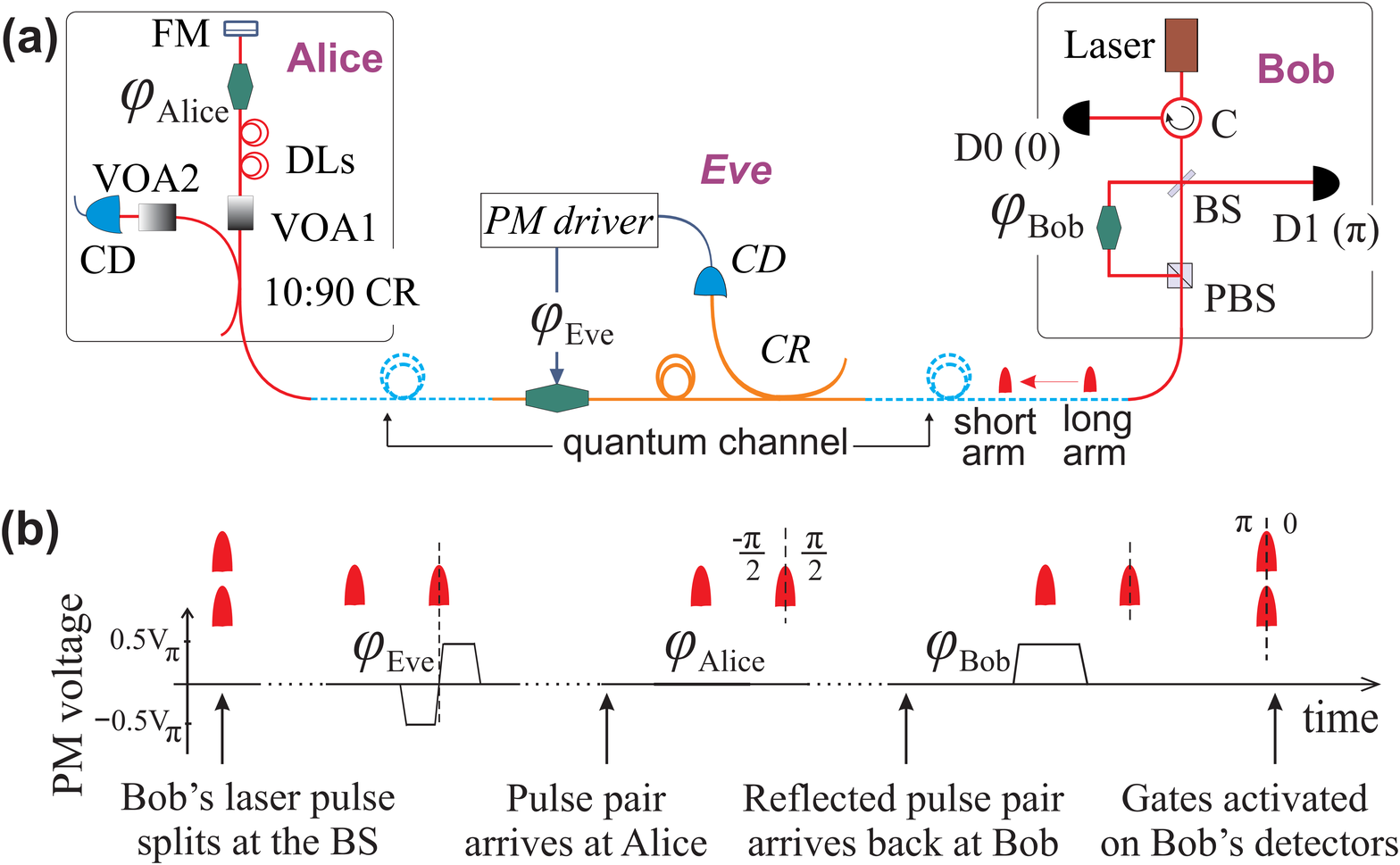}}
\caption{Manipulation of the calibration routine: \textbf{(a)} Simplified version of Alice and Bob devices and Eve (in italic) gearing for the hack. FM: Faraday mirror, CD: classical photodiode, DLs: delay loops, VOA: variable optical attenuator, CR: coupler, BS: 50:50 beam splitter, PBS: polarizing beam splitter, C: optical circulator. The hexagonal-shaped objects are phase modulators (PMs); $\varphi_{\rm X}$, where X is Bob, Alice or Eve, represents the applied modulation. \textbf{(b)} Timeline for a cycle of the hacked LLM. $V_{\pi}$: PM voltage for a $\pi$ phase shift.}
\label{fscheme}
\end{figure}

The optical setup of Clavis2 is based on the plug-and-play QKD scheme~\cite{clavis2guide,pnp}. An asymmetric Mach-Zehnder interferometer operates in a double pass over the quantum channel by using a Faraday mirror; see Fig.~\ref{fscheme}(a) without Eve. The interference of the paths taken by two pulses travelling from Bob to Alice and back is determined by their relative phase modulation ($\varphi_{\rm Bob} - \varphi_{\rm Alice}$), and forms the principle for encoding the key. Any birefringence effects of the quantum channel are passively compensated. As a prerequisite to the key exchange, Clavis2 calibrates its detectors in time via a sequence named Line Length Measurement (LLM). Bob emits a pair of \emph{bright} pulses and applies a series of detector gates around an initial estimate of their return. The timing of the gates is electronically scanned (while monitoring detector clicks) to refine the estimation of the channel length and relative delay between the time of arrival of the pulses at D0 and D1. Alice keeps her phase modulator (PM) switched off, while Bob applies a uniform phase of $\pi/2$ to one of the incoming pulses. Therefore, both detectors are equally illuminated and their detection efficiencies, denoted by $\eta_0(t)$ and $\eta_1(t)$, can be resolved in time. Any existing mismatch can thus be minimized by changing the gate-activation times (see Fig.~\ref{dsb}). 

However, the calibration routine does not always succeed; as reported in~\cite{zhao08}, a high detector efficiency mismatch (DEM) is sometimes observed after a normal run of LLM. For example, we have noticed a temporal mismatch as high as $400$ ps in Clavis2. This physical limitation of the system -- arising due to fast and uncontrollable fluctuations in the quantum channel or electromagnetic interference in the detection circuits -- is the vulnerability that the TSA exploits. However, the attack has some limitations: it is applicable only when the temporal mismatch happens to exceed a certain threshold value, which is merely $4\%$ of all the instances~\cite{zhao08}. Also, Eve can neither control the mismatch (as it occurs probabilistically), nor extract its value (as it is not revealed publicly).

We exploit a weakness of the calibration routine to induce a large and deterministic DEM without needing to extract any information from Bob. As depicted in Fig.~\ref{fscheme}(a), Eve installs her equipment in the quantum channel such that the laser pulse pair coming out of Bob's short and long arm passes through her PM. Eve's modulation pattern is such that a rising edge in the PM voltage flips the phase in the second (long arm) optical pulse from $-\pi/2$ to $\pi/2$, as shown in Fig.~\ref{fscheme}(b). As a result of this hack, when the pulse pair interferes at Bob's 50:50 beam splitter, the two temporal halves have a relative phase difference ($\varphi_{\rm Bob} - \varphi_{\rm Eve}$) of $\pi$ and $0$, respectively. This implies that photons from the first (second) half of the interfering pulses yield clicks in D1 (D0) deterministically. As the LLM localizes the detection efficiency peak corresponding to the optical power peak, an \emph{artificial} temporal displacement in the detector efficiencies is induced. An inverse displacement can be obtained by simply inverting the polarity of Eve's phase modulation.  

\begin{figure}\tt
\centerline{\includegraphics[width=8.6cm]{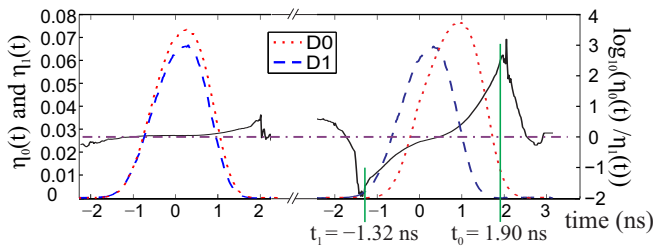}}
\caption{Induced temporal mismatch: Efficiencies $\eta_0(t)$ (dotted) and $\eta_1(t)$ (dashed) from normal LLMs, on the left, and after Eve's hack that induced a separation of $459$ ps, on the right. The logarithm of their ratio, quantifying the degree of mismatch (solid line), is at least an order of magnitude higher in the flanks after Eve's hack: the dash-dot line indicates zero mismatch. To eavesdrop successfully, Eve times the arrival of ``appropriately bright'' faked states at $t=t_0$ or $t_1$ in Bob.}
\label{fceff}
\end{figure}

In the supplementary section~\cite{suppref}, we describe a proof-of-principle experiment to deceive the calibration routine. With this setup, we record the temporal separation $\Delta_{01}$, i.e.\ the difference between the delays for electronically gating D0 and D1, for several runs of LLM. Relative to the statistics from the normal runs (denoted by $\Delta^{\rm{no \, Eve}}_{01}$), the hacked runs yield an average shift, $\Delta^{\rm{Eve}}_{01} - \Delta^{\rm{no \, Eve}}_{01}$ = $459$ ps with a standard deviation of $105$ ps. Figure~\ref{fceff} shows the detection efficiencies $\eta_0(t)$ and $\eta_1(t)$ (measurement method explained in~\cite{suppref}) for the normal and hacked cases. It also provides a quantitative comparison between the usual and induced mismatch. Note that a larger mismatch can be obtained by modifying the shape of laser pulses coming from Bob.

After inducing this substantial efficiency mismatch, Eve can use an intercept-resend strategy employing `faked states'~\cite{makarov05} to impose her will upon Bob (and Alice). Compared to her intercepted measurements, she prepares the opposite bit value in the opposite basis and sends it with such a timing that the detection of the opposite bit value is suppressed due to negligible detection efficiency. As an example, assume that Eve measures bit $0$ in the $Z$ basis [in a phase-coded scheme, measuring in $Z$ $(X)$ basis $\Leftrightarrow\text{applying } \varphi=0 \,\, \left(\pi/2\right)$]. Then, she resends bit $1$ in the $X$ basis, timed to be detected at $t = t_0$ (see Fig.~\ref{fceff}), where D1 is almost blind. Using the numerical data on the induced mismatch, Eq.~$3$ from~\cite{makarov06} yields a QBER $<0.5\%$ if the FSA is launched at times $t_0$ and $t_1$ where the efficiency mismatch is high.

However, it can be observed that the detection probabilities for D0 and D1 are quite low in this case. A considerable decrease in the rate of detection events in Bob could ensue an alarm. Also, the (relatively increased) dark counts would add significantly to the QBER. In fact, Eve needs to \emph{match} the channel transmission $T$ that Alice and Bob expect, without exceeding the QBER threshold at which they abort key generation~\cite{gllp}. Experimentally, we find that the abort threshold depends on the channel loss seen by Clavis2; for an optical loss of $1\:\!$--$\:\!6 \:\, \text{dB}$ (corresponding to $0.79\!\;\!>\!\!\;T\;\!\!>\;\!\!0.25$), it lies between $5.94\:\!$--$\:\!8.26\%$. 

\begin{table}
\newcommand\Ta{\rule{0pt}{2.9ex}}
\newcommand\Tb{\rule{0pt}{2.4ex}}
\begin{tabular}{|c|c|c|c|}
\hline
$\rightarrow$Eve \Tb & Eve$\rightarrow$ & Bob's result & Detection probability \\
\hline
$Z,0$ \Tb & $\, X,1,\mu_0,t_0 \,$ & $0$ & $\mathbf{q}_0 = d_0 + \left(1 - d_0\right)\times$  \\
 & & & $\left(1 - \text{exp}\left(- \mu_0 \eta_0(t_0)/2\right)\right)$ \\ \cline{3-2} \cline{4-2}
 \Tb & & $1$ & $\mathbf{q}_1 = d_1 + \left(1 - d_1\right)\times$  \\
 & & & $\left(1 - \text{exp}\left(- \mu_0 \eta_1(t_0)/2\right)\right)$ \\ \cline{3-2} \cline{4-2}
 \Tb & & $0 \cap 1$ & $\mathbf{q}_0\mathbf{q}_1$  \\ \cline{3-2} \cline{4-2}
 \Tb & & loss & $1 - \left(\mathbf{q}_0 + \mathbf{q}_1 - \mathbf{q}_0\mathbf{q}_1\right)$ \\ 
\hline
$X,0$ \Tb & $\, Z,1,\mu_0,t_0 \,$ & $0$ & $\mathbf{r}_0 = d_0$ \\ \cline{3-2} \cline{4-2}
 \Tb & & $1$ & $\mathbf{r}_1 = d_1 + \left(1 - d_1\right)\times$  \\
 & & & $\left(1 - \text{exp}\left(- \mu_0 \eta_1(t_0)\right)\right)$ \\ \cline{3-2} \cline{4-2}
 \Tb & & $0 \cap 1$ & $\mathbf{r}_0\mathbf{r}_1$  \\ \cline{3-2} \cline{4-2}
 \Tb & & loss & $1 - \left(\mathbf{r}_0 + \mathbf{r}_1 - \mathbf{r}_0\mathbf{r}_1\right)$ \\
\hline
$X,1$ \Tb & $\, Z,0,\mu_1,t_1 \,$ & $0$ & $\mathbf{s}_0 = d_0 + \left(1 - d_0\right)\times$  \\
 & & & $\left(1 - \text{exp}\left(- \mu_1 \eta_0(t_1)\right)\right)$ \\ \cline{3-2} \cline{4-2}
 \Tb & & $1$ & $\mathbf{s}_1 = d_1$ \\ \cline{3-2} \cline{4-2} 
 \Tb & & $0 \cap 1$ & $\mathbf{s}_0\mathbf{s}_1$  \\ \cline{3-2} \cline{4-2}
 \Tb & & loss & $1 - \left(\mathbf{s}_0 + \mathbf{s}_1 - \mathbf{s}_0\mathbf{s}_1\right)$ \\
\hline 
\end{tabular} 
\caption{Faked-state attack, given that Alice prepared bit $0$ in the $Z$ basis and that Bob measured in the $Z$ basis (only matching basis at Alice and Bob remains after sifting). The first column contains the basis chosen by Eve and her measurement result. The second column shows parameters of the faked state resent by Eve: basis, bit, mean photon number, timing. The third column shows Bob's measurement result; $0\cap1$ denotes a double click. The last column shows the corresponding click probabilities (ignoring possible superlinearity effect in gated detectors~\cite{lydersen11}). Note: The first result $\left(\rightarrow\text{Eve} \equiv Z,0\right)$ is twice as likely to occur as the other two.}
\label{table:attack}
\end{table} 

Eve solves these problems by increasing the mean photon number of her faked states. To evaluate her QBER, we elaborate the approach of~\cite{makarov06} by generalizing table~I from this reference. Our attack strategy, carefully accounting for all the involved factors, is summarized in Table~\ref{table:attack}. For instance, in the first row we replace the probability of detection $\eta_0(t_0)/2$ by $1 - \text{exp}\left(- \mu_0 \eta_0(t_0)/2\right)$ for a coherent-state pulse of mean photon number $\mu_0$ impinging on Bob's detectors at time $t_0$. Including the effect of the dark counts into this expression, Bob's probability to register $0$ becomes $\mathbf{q}_0 = d_0 + \left(1 - d_0\right)\left(1 - \text{exp}\left(- \mu_0 \eta_0(t_0)/2\right)\right)$, where $d_{0}$ is the dark count probability in detector D0. A row for double clicks, i.e.\ simultaneous detection events in D0 and D1, is added for every (re-sent) state.

Due to the FSA, the D0/1 click probability at time $t$ no longer depends solely upon $\eta_{0/1}(t)$. Summing over all the states sent by Alice (by extending Table~\ref{table:attack}), the total detection probabilities in D0 and D1 when the attack is launched at specific times $t_0$ and $t_1$ are
\begin{align}
p&_0(\mu_0,\mu_1) = 0.75+0.25d-0.25(1-d)\times \nonumber \\
& (e^{-0.5\mu_0\eta_{00}} + e^{-0.5\mu_1\eta_{01}} + e^{-\mu_1\eta_{01}})\, , \\
p&_1(\mu_0,\mu_1) = 0.75+0.25d-0.25(1-d)\times \nonumber \\
& (e^{-0.5\mu_0\eta_{10}} + e^{-0.5\mu_1\eta_{11}} + e^{-\mu_0\eta_{10}})\, .
\label{p0p1}
\end{align}
Here $\eta_{jk} = \eta_j(t_k)$ with $j,k \in \{0,1\}$ and $d = \text{mean}\left(d_0,d_1\right)$ are used to simplify the expressions. Similarly, one can compute the expression for $p_{0\cap1}$, the total double-click probability. Eve's error probability, the arrival probability of the optical signals in Bob, and the QBER are
\begin{align}
p&_{\rm error}(\mu_0,\mu_1) = 0.75+0.25d-0.5p_{0\cap1}-0.125\times \\
& (1-d)\left(e^{-\mu_0\eta_{10}}+2e^{-0.5\mu_0\eta_{10}}+e^{-\mu_1\eta_{01}}+2e^{-0.5\mu_1\eta_{01}}\right) , \nonumber \\
p&_{\rm arrive}(\mu_0,\mu_1) = p_{0} + p_{1} - p_{0\cap1}\, , \\
\text{Q}&\text{BER}(\mu_0,\mu_1) = p_{\text{error}}(\mu_0,\mu_1)/p_{\text{arrive}}(\mu_0,\mu_1)\, .
\end{align}
Here double clicks are assumed to be assigned a random bit value by Bob~\cite{dblclk}, causing an error in half the cases. 

\begin{figure}\tt
\centerline{\includegraphics[width=6.5cm]{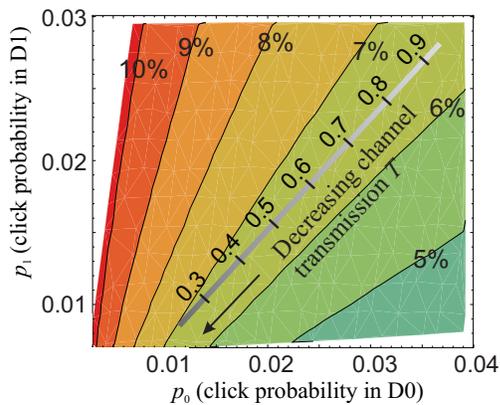}}
\caption{Minimum QBER versus click probabilities in D0 and D1: Eve minimizes the error with a suitable choice of the mean photon number of the faked states (for this plot, $1<\mu_0<100$ and $21<\mu_1<120$ at Bob's detectors). The thick shaded line indicates Bob's detection probabilities. The QBER introduced by Eve stays below 7\% for $T \gtrsim 0.25$.}
\label{ftmshftNqber}
\end{figure}

If Alice and Bob are connected back-to-back (channel transmission $T \approx 1$), the click probabilities in Bob should be slightly less than half of the peak values in Fig.~\ref{fceff}. This is owing to optical losses ($\gtrsim 3\,\text{dB}$) in Bob's apparatus. Eve's constraints can now be formalized as: starting in the vicinity of $p_0=0.038$ and $p_1=0.032$, not only does she have to match Bob's expected detection rate for any given $T<1$, but also keep the resultant QBER below the threshold at which Clavis2 aborts the key exchange. We assume Eve detects photons at Alice's exit using a perfect apparatus, and resends perfectly aligned faked states.

Substituting $t_1 = -1.32$ ns, $t_0 = 1.90$ ns (marked in Fig.~\ref{fceff}) and $d = 2.4 \times 10^{-4}$ in Eqns.\ 1--5, Eve collects tuples $\left[p_0, \, p_1, \: \text{QBER} \right]$ by varying $\mu_0$ and $\mu_1$ in a suitable range. Out of all tuples that feature the same detection probabilities (arising from different combinations of $\mu_0$ and $\mu_1$), Eve chooses the one having the lowest QBER. A contour plot in Fig.~\ref{ftmshftNqber} displays this minimized error $\min_{\mu_0,\mu_1}\text{QBER}\left((\mu_0,\mu_1)\vert\left(p_0,p_1\right)\right)$. The thick shaded line shows that for $T>0.25$, Eve not only maintains the detection rates within $5\%$ of Bob's expected values, but also keeps the QBER below 7\% \footnote{The QBER can be reduced even further if Bob checks only the \emph{overall} detection probability $p_0 + p_1$.}; thus breaking the security of the system. Note that the simulation assumes a lossless Eve, but in principle she can cover loss from her realistic detection apparatus by increasing $\mu_0$ and $\mu_1$ further and/or including $t_0$ and $t_1$ in the minimization.
 
To counter this hack, Bob should randomly apply a phase of $0$ or $\pi$ (instead of $\pi/2$ uniformly) while performing LLM. This modification is implementable in software and has already been proposed to ID~Quantique. More generally, a method to shield QKD systems from attacks that exploit DEM is described in Ref.~\cite{larspra11}. 

In conclusion, we report a proof-of-principle experiment to induce a large detector efficiency mismatch in a commercial QKD system by deceiving a vital calibration routine. An optimized faked-state attack on such a compromised system would not alarm Alice and Bob as it would introduce a QBER $< 7\%$ for a large range of expected channel transmissions. Thus, the overall security of the system is broken. With initiatives for standardizing QKD \cite{qkdstdzn} underway, we believe this report is timely and shall facilitate elevating the security of practical QKD systems.

\textbf{Acknowledgments}:
We thank M.\ Legr\'e from ID~Quantique and N.\ L\"utkenhaus for helpful discussions; Q.\ Liu, L.\ Meier and A.\ K\"appel for technical assistance. This work was supported by the Research Council of Norway (grant no.\ 180439/V30), DAADppp mobility program financed by NFR (project no.\ 199854) and DAAD (project no.\ 50727598), and BMBF CHIST-ERA (project HIPERCOM). Ca.~Wi.\ acknowledges support from FONCICYT project no.\ 94142.

\appendix
\onecolumngrid
\section{\large{D\MakeLowercase{evice calibration impacts security of quantum key distribution}: T\MakeLowercase{echnical appendix}}}
\begin{figure}[h]
\centerline{\includegraphics[width=8.4cm]{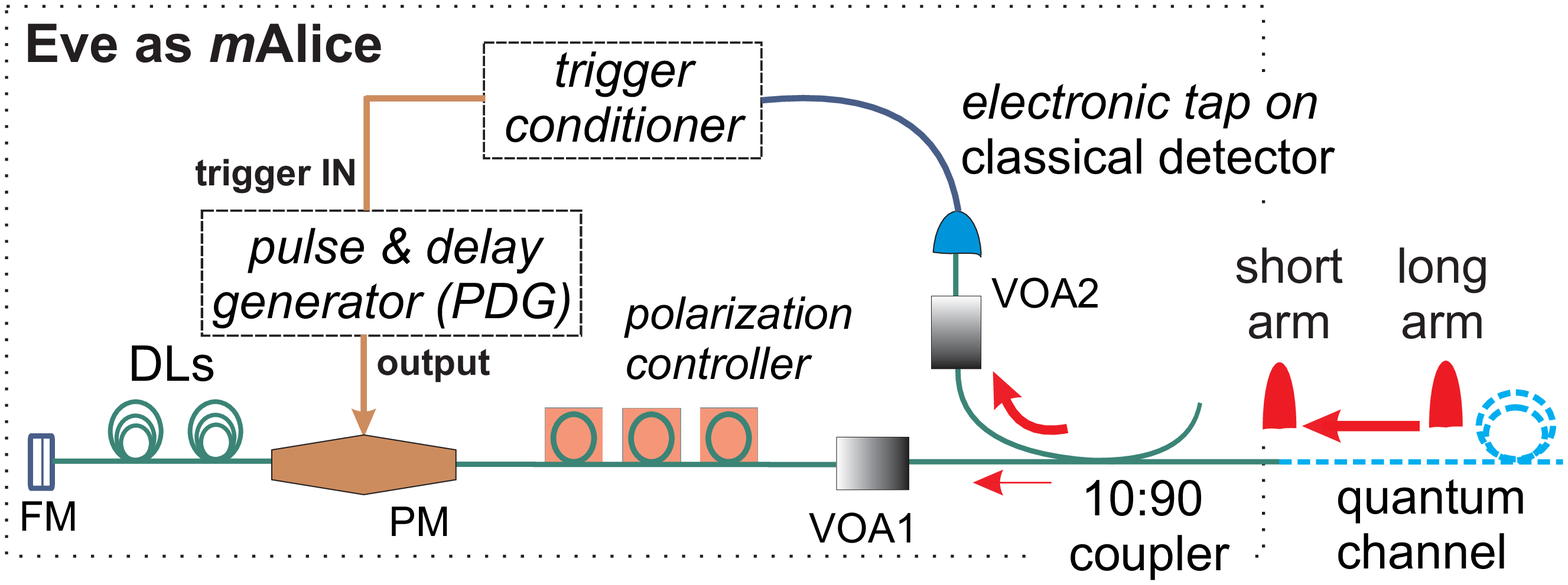}}
\caption{Eve's implementation ($m$Alice) by modifying Alice's module: The onboard pulser driving the phase modulator (PM) is disconnected, and the PM itself is positioned \emph{before} the 23.5~km delay loops (DLs). The trigger conditioner circuit allows (prevents) the pulse \& delay generator to be triggered by the short arm (long arm) optical pulses. Newly added components to the original Alice module are labeled in italic. VOA: variable optical attenuator, FM: Faraday mirror.}
\label{fimplmntn}
\end{figure}
\noindent \textbf{Implementation of the hack:}
Here, we explain our experimental implementation of the scheme outlined in the Letter for deceiving Line Length Measurement (LLM), the calibration routine of the Clavis2 QKD system~\cite{clavis2guide}. For this purpose, we rig the module of Alice as shown in Fig.~\ref{fimplmntn}. From now on, we call this manipulated device $m$Alice. An electronic tap placed on the classical detector (normally used by Alice for measuring the incoming optical power~\cite{trojanH}) is conditioned appropriately with a homemade circuit. The output of this circuit provides the trigger for the pulse \& delay generator (PDG, Highland Technology P400), which essentially drives the phase modulator (PM) in $m$Alice.

For experimental convenience, we also change the settings in the Clavis2 firmware (Bob's EEPROM specifically) such that during the execution of LLM, $\varphi_{\rm Bob} = 0$ is applied instead of the usual $\pi/2$. This relaxes the requirement on Eve's modulation pattern: in comparison to the waveform in Fig.~2(b) in the Letter, the PDG needs to switch simply from $0$ to $V_{\pi}$ through the center of the optical pulse. This is in principle equivalent to the scheme in Fig.~2(b) in the Letter, while easier to implement. In other words, it does not affect a full implementation of Eve. Normally, Alice applies the phase modulation in a double pass by making use of the Faraday mirror. However, the PM in $m$Alice is shifted closer to Alice's entrance (i.e.\ before the delay loops) to enable a precise synchronization of the PDG. To ensure that the photons passing through the PM (in a single pass now) pick up the requisite `$\pi$' modulation, a polarization controller is deployed before the PM. 

Finally, the synchronization of the rising edge of Eve's modulation to the center of the optical pulse is performed by scanning the delay in the PDG (in steps of 5~ps) while monitoring the interference visibility \cite{clavis2guide}. As Eve's modulation flips the phase of the optical pulse through the center, the visibility reduces to zero. The corresponding delay setting of the PDG can then be used to induce the temporal efficiency mismatch between Bob's detectors D0 and D1, during the execution of LLM.

We emphasize that the $m$Alice module serves as a proof-of-principle implementation \emph{only} for inducing the detector efficiency mismatch during the LLM. It should not be confused with Eve's intercept or resend modules, needed in the subsequent faked-state attack. Finally, note that Eve is free to modify Bob's pulses or replace them by her suitably-prepared pulses, and thus effectively control the amount of detection efficiency mismatch that can be induced.\\

\vspace{2.0mm}

\noindent \textbf{Measurement of efficiency curves:}
Detection efficiencies $\eta_0(t)$ and $\eta_1(t)$ are estimated at single-photon level by scanning the detector gates in steps of 20~ps with an external laser (optical pulse-width $\sim 200$~ps). We average the click probability per gate and subtract $d_{0/1}$ (the dark count rate in D0/1) from it. This gives a more accurate estimate of the efficiencies, especially in the flanks (see Fig.~3 in the Letter). 

\end{document}